\documentclass[aps,prb,twocolumn]{revtex4-2} %new standard
\newcommand{\beq}{\begin{equation}}
\newcommand{\beqn}{\begin{eqnarray}}
\newcommand{\eeq}{\end{equation}}
\newcommand{\eeqn}{\end{eqnarray}}

\newcommand{\ket}[1]{\vert #1 \rangle}

\newcommand{\eV}{{\text{eV}}}
\makeatletter

\newcommand{\Rmnum}[1]{\expandafter\@slowromancap\romannumeral #1@}
\makeatother

\usepackage{graphics,graphicx,amsmath,hyperref}
\usepackage{tabularx,float}
\usepackage{epsfig}
\usepackage{subfigure}
\usepackage{bm}
\usepackage{wasysym}
\usepackage{bm}

\usepackage{color}
\usepackage{verbatim}
\graphicspath{{fig/}}

\begin{document}

\date{\today}

\author{X. Lu}
\affiliation{Laboratoire de Physique des Solides, Universit\'e Paris Saclay, CNRS UMR 8502, F-91405 Orsay Cedex, France}

\author{M. O. Goerbig}
\affiliation{Laboratoire de Physique des Solides, Universit\'e Paris Saclay, CNRS UMR 8502, F-91405 Orsay Cedex, France}

\title{Dirac quantum well engineering on the surface of topological insulator}

\begin{abstract}

We investigate a quantum well that consists of a thin topological insulator sandwiched between two trivial insulators. More specifically,
we consider smooth interfaces between these different types of materials such that the interfaces host not only the chiral interface states,
whose existence is dictated by the bulk-edge correspondence, but also massive Volkov-Pankratov states. We investigate possible hybridization between 
these interface states as a function of the width of the topological material and of the characteristic interface size. Most saliently, we find a 
strong qualitative difference between an extremely weak effect on the chiral interface states and a more common hybridization
of the massive Volkov-Pankratov states that can be easily understood in terms of quantum tunneling in the framework of the model of a (Dirac) quantum well we introduce here.

\end{abstract}

%\pacs{78.30.Na, 73.43.Lp, 81.05.Uw}

\maketitle

\section{Introduction}

Topological insulators \cite{kane2010rmp,qixiaoliang2011rmp} (TIs) are insulating materials that exhibit chiral conducting surface states.
This exotic property is a manifestation of the bulk-edge correspondence which states that the topological invariant of the bulk 
Hamiltonian dictates the presence of gapless chiral edge or surface states. The latter have been observed experimentally by angle-resolved photo-emission 
spectroscopy \cite{bianchi2010coexistence,bianchi2011prl,chenchaoyu2012pnas} or via quantized conductances, e.g., in HgTe/CdTe quantum wells \cite{BHZ2006science,Konig2007science}. 

Already in the 1980s, two decades before the advent of topological materials in general and TIs in particular, the bulk-edge correspondence
had been theorized in inverted-gap systems \cite{volkov1985two,pankratov1987supersymmetry,Fradkin86,Fradkin87}, most prominently by Volkov and Pankratov in the context of HgTe/CdTe 
heterostructures \cite{volkov1985two,pankratov1987supersymmetry}. Indeed, they found that HgTe has an inverted gap in the 
band structure as compared to CdTe. In the modern language of topological band theory, this is precisely a consequence of a difference in the bulk invariant characterizing the two 
materials \cite{BHZ2006science}. At the interface of a HgTe/CdTe heterostructure, the gap therefore needs to change sign, and a robust, topologically protected,
chiral state thus emerges. Moreover, Volkov and Pankratov showed in their seminal work that, in the case of a smooth change of the gap parameter over the interface,
massive surface states, now called \textit{Volkov-Pankratov} (VP) states, can occur beyond the chiral ones. Unlike in Ref. \cite{mahler2019}, VP states mean here only the massive surface states and we do not call the chiral states the massless VP states.
Only recently, these massive surface states have regained interest, namely due to their experimental observation in transport measurements in
HgTe/CdHgTe heterojunctions \cite{inhofer2017observation}. Theoretical studies by Tchoumakov \textit{et al.} showed that the occurrence of such states is generic in what is
now called a \textit{topological heterojunction} \cite{tchoumakov2017volkov}, i.e., a smooth interface between a topological material and a trivial insulator. 
Indeed, they have been shown to arise not only in TIs \cite{tchoumakov2017volkov,inhofer2017observation,mahler2019,vdBerg2020}, but also in interfaces of Weyl 
semimetals \cite{grushin2016,tchoumakov2017weyl,mukherjee2019dynamical}, topological graphene nanoribbons \cite{berg2020volkov} and topological superconductors \cite{alspaugh2020volkov}.
Furthermore, the magneto-optical properties of smooth topological interfaces have been studied both in the context of TIs \cite{Lu_2019,vdBerg2020} 
and Weyl semimetals \cite{mukherjee2019dynamical} in the prospect of a, to the best of our knowledge yet missing, direct spectroscopic identification of massive VP states.

\begin{figure}[h]
    \centering
    \includegraphics[width=0.4\textwidth]{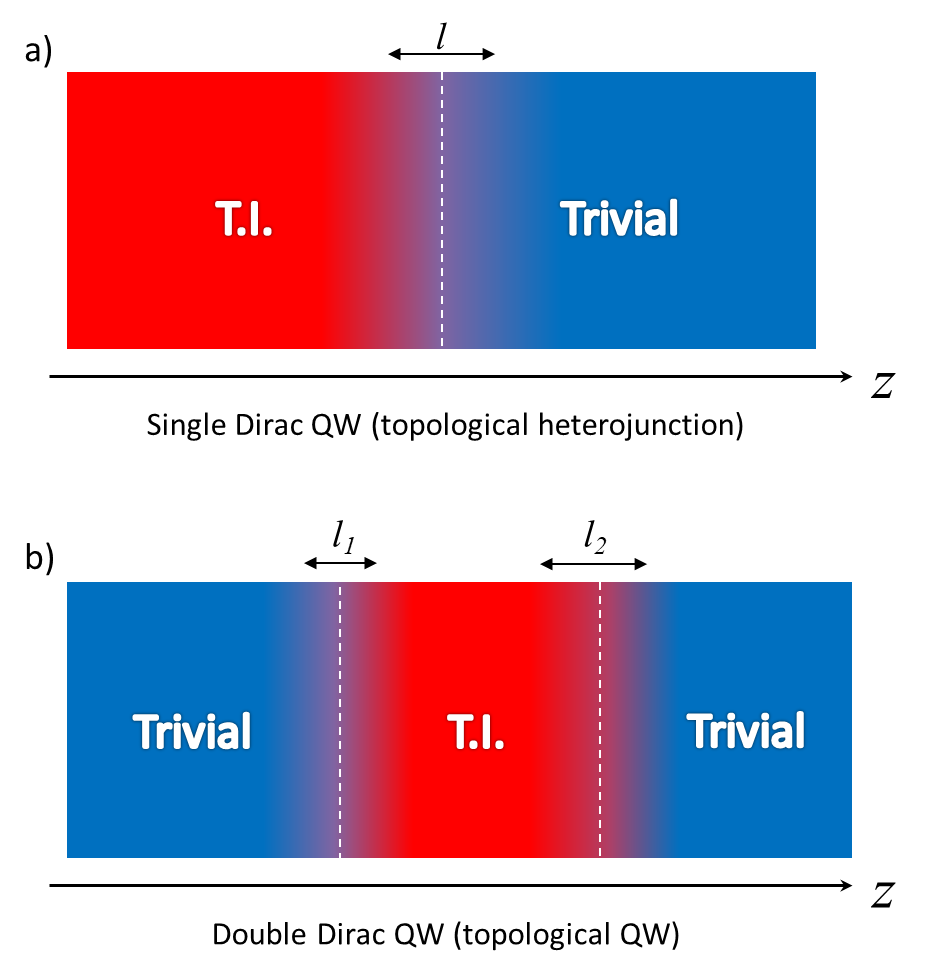}
    \caption{(a) Schematic of a single Dirac QW. (b) Schematic of a double Dirac QW. The spatially varying gap $\Delta(z)$ switches from positive sign to negative when one passes from the blue area (trivial phase) to the red area (topological phase) and the gap is vanishing somewhere in between. $l$, $l_1$ and $l_2$ characterize the smoothness of domain wall between phases.}
    \label{fig:diracqw}
\end{figure}

Previous theoretical studies interpreted these emergent massive states as either another type of solutions of differential 
equation \cite{volkov1985two,pankratov1987supersymmetry} or Landau quantization induced by a pseudomagnetic field, i.e.,
the smoothness \cite{grushin2016,tchoumakov2017volkov,Roy_2018,Lu_2019,ilan2020}. In the present paper, we adopt a complementary perspective on VP states and topological chiral states in the framework of quantum well physics. Indeed, the matrix model -- it is at least a two-band model -- which describes the topological heterojunction, 
can be transformed, within supersymmetric quantum mechanics \cite{landwehr2012landau}, in such a manner that the components of the wavefunction satisfy a more conventional 
Schr\"odinger equation in a modified well potential \cite{WITTEN1981513,pankratov1987supersymmetry} that arises from the (linearly) varying gap parameter. 
One is therefore confronted effectively with the conventional problem of a onedimensional (1D) quantum mechanical particle in a quantum well, which we call henceforth
\textit{Dirac quantum well} (QW), as a complementary and equivalent point of view with respect to the topological heterojunction. As shown in Fig. \ref{fig:diracqw}, 
a single interface in the form of a topological heterojunction thus gives rise to a single Dirac QW within this treatment, which we will review in detail in Sec. \ref{sec:2}.
More importantly for the present work, a thin TI sandwiched between two trivial insulators, such as in a CdTe/HgTe/CdTe heterostructure that is commonly said to be a \textit{single} QW, can be viewed as 
a \textit{double} Dirac QW. 

This complementary framework, i.e., the description of topological heterojunctions in terms of Dirac QWs, has two major advantages. 
The first one is conceptional: one can understand explicitly within the Dirac QW the appearance of VP states in a smooth topological 
heterojunction in terms of a quantum confinement effect. Indeed, one finds always at least one bound state in one of the chiral sectors for which the potential is necessarily 
confining, and this state corresponds to the robust chiral surface state. Furthermore, a smoother interface yields, somewhat unexpectedly, a \textit{shallower confinement} 
with more bound solutions in the Dirac QW for both chiralities. The second advantage is practical. Most studies 
on VP states have been restricted so far to a single boundary problem, i.e., the possible
coupling between the surface states located at two opposite surfaces of a TI with finite width has not been considered. However, many theoretical and experimental 
papers \cite{lu2010massive,liu2010oscillatory,linder2009anomalous,neupane2014observation,DiazFernandez2017} are interested in quantum tunneling between the chiral states on two sides of a 
thin TI film. The Dirac QW turns out to be a suitable framework to discuss both the tunneling between the chiral states and that between VP states which reside at different sides of a finite system.

The main findings of our work are the following. We show that the topological chiral state is actually the ground state of Dirac QW. One can engineer the depth 
and the width of Dirac QW by changing the smoothness of the interface. Furthermore, we also study quantum tunneling between two adjacent Dirac QWs that are inherently asymmetric. 
A plausible realization of two Dirac QWs would be a finite-sized TI. For the massive states, the quantum tunneling strength depends on the smoothness and the distance of two Dirac QWs in a similar 
way as in the case of quantum tunneling between two conventional QWs separated by a potential barrier of finite width and height. The behavior of the massless chiral states
is strikingly different: within the Dirac QW model, we find indeed a shift away from zero energy that leads to a small mass gap in these states. However, there is no direct 
hybridization in the absence of a perturbation that couples the two different chiralities, and the mass gap is found to be several orders of magnitude smaller than the direct
hybridization of the massive VP states. This particular feature provides a complementary quantum-mechanical view on 
topological protection of chiral surface states.

The paper is organized in the following manner. In Sec. \ref{sec:2}, we provide an introduction to the link between a topological heterojunction and the Dirac QW. We review here
the existence, in an explicit as well as a general treatment, of a chiral state and show how the emergence of the massive surface states can be understood as a quantum confinement 
effect in terms of Dirac QWs. Sec. \ref{sec:3} presents our main results on the tunneling effect between massive VP states, which we discuss in the framework of 
two coupled Dirac QWs, as a function of the interface smoothness and their relative separation. In particular, we give an analytical recipe to estimate the energy splitting of 
the massive VP states in comparison with the chiral ones. We also show the similarities and the peculiarities of a double Dirac QW when compared to conventional a double square QW.

\section{Quantum confinement: Dirac quantum well} \label{sec:2}

In this section, we introduce the concept of Dirac QW via an explicit example. We consider here a single boundary between a threedimensional (3D) TI and trivial insulator modeled by a 3D TI Hamiltonian  
with spatially varying gap parameter. The caveat behind this model is the following: we consider that the low-energy model in the vicinity of a topological phase transition
can be described in terms of a massive Dirac fermion the gap parameter (mass) of which is, say, negative in the topologically nontrivial phase and positive otherwise.  
Suppose that the half-space $z<0$ is filled by a topological phase and the other half-space $z>0$ filled by a trivial one [see Fig. \ref{fig:diracqw}(a)]. 
The situation can be described with the generic Hamiltonian \cite{zhang2012}
\begin{equation}
    \label{eq:h03dti}
    H_0 = \Delta (z) \tau_z + \hbar v k_z \tau_y + \hbar v \tau_x (k_y \sigma_x - k_x \sigma_y)
\end{equation}
for a massive Dirac fermion,
where $v$ is Fermi velocity and $\Delta(z)$ is the gap parameter (half of the spatially varying gap). While the Pauli matrices $\sigma_\mu$ represent here the true spin of the system with
underlying spin-orbit coupling, the Pauli matrices $\tau_\mu$ represent another (lattice) degree of freedom, such as for example orbitals in a multi-orbital system as mentioned 
above.
The gap parameter $\Delta(z)$ changes its sign across the interface. By construction, one enters into a topological phase (with an \textit{inverted} gap) 
when $\Delta(z)<0$ and into a trivial phase when $\Delta(z)>0$. For simplicity, we suppose additionally $\Delta(z)$ to be an increasing function of $z$ and:
    \begin{gather}
    	\Delta(z) = 
        \begin{cases}
        	-\Delta_0 & \text{if $z \to -\infty$} \\
            \Delta_0 &  \text{if $z \to +\infty$}
        \end{cases}
        \label{exp:deltaz}
    \end{gather}
where $\Delta_0 > 0$ is half of the bulk gap. Here, $k_z$ should be replaced by $-i \partial_z$ because of its non-commutativity with $\Delta(z)$ while the components 
of the wavevector in the interface plane remain good quantum numbers. Even without the explicit calculation of the spectrum of Hamiltonian (\ref{eq:h03dti}), we may already
appreciate an important point here. Due to the spatial variation of the gap function $\Delta(z)$, the electronic motion in the $z$-direction is generically quantized into 
$(d-1)$-dimensional surface bands if we start from a $d$-dimensional bulk system. While we illustrate our model in 3D heterostructures, it is thus directly 
applicable to other spatial dimensions such as 2D TIs. It is more convenient to work in the \textit{Weyl basis}, which is obtained 
by the unitary transformation $T = \exp( i \pi \tau_y / 4)$ that simply interchanges the role of $\tau_x$ and $\tau_z$ ($\tau_x\rightarrow \tau_z$ and $\tau_z\rightarrow -\tau_x$),
and then solve the equation $H_T \ket{\psi} =  E \ket{\psi}$ where $H_T = T H_0 T^{\dagger}$. In this basis, the Hilbert space can be 
decomposed into an orthogonal direct sum of two subspaces with opposite chiralities. Notice that the eigenstates $\ket{\psi}$ are four-component spinors that can be written as
    \begin{align}
        \ket{\psi} =
        \begin{pmatrix}
            \chi_{+} (z) \\
            \chi_{-} (z)
        \end{pmatrix},
    \end{align}
where $\chi_{\pm}$ are themselves two-component spinors of chirality $\pm$. We obtain thus a set of two differential equations \cite{volkov1985two,pankratov1987supersymmetry}:
\begin{equation}\label{eq:schrodinger0}
     \left(E^2 - \hbar^2 v^2 k_{\parallel}^2 \right) \chi_{\lambda} = \left[\Delta(z)+\lambda\hbar v\partial_z\right]\left[\Delta(z)-\lambda\hbar v\partial_z\right]\chi_\lambda,
\end{equation}
where $k_{\parallel}^2 = k_x^2 + k_y^2$, and $\lambda = \pm$ represents the chirality. 

Let us now consider the differential equations (\ref{eq:schrodinger0}) for the two chiral sectors in terms of a 1D quantum mechanical problem. 
Indeed, the equations can be rewritten as
    \begin{equation}
    \label{eq:schrodinger}
    \left(E^2 - \hbar^2 v^2 k_{\parallel}^2 \right) \chi_{\lambda} = \Tilde{E}_\lambda^2\chi_\lambda
    = \left(-\hbar^2 v^2 \partial_z^2 + U_{\lambda} (z) \right) \chi_{\lambda},
\end{equation}
the right hand side of which shows now a second-order derivative in $z$, as it is the case for a 1D Schr\"odinger equation with a confining potential 
\begin{equation}\label{eq:superpot}
U_{\lambda}(z) = \Delta(z)^2 + \lambda \hbar v \partial_z \Delta(z)
\end{equation}
which 
itself depends on the chirality $\lambda$. Solving $E$ for the Hamiltonian $H_T$ is equivalent to solving 
\begin{equation}\label{eq:energies}
\Tilde{E}_\lambda^2 \equiv E^2 - \hbar^2 v^2 k_{\parallel}^2
\end{equation}
for this Schr\"{o}dinger equation whose spectrum $\Tilde{E}_\lambda^2$ must be non-negative. Note that the spectrum and the potential in the Schr\"odinger-type equation \eqref{eq:schrodinger}, which we have just obtained, have the physical dimension of a squared
energy. To emphasize that we are working with such auxiliary quantities that do not have the dimension of energy (but its square), we explicitly use, in the following paragraphs,
the term \textit{virtual energies} when we consider the context of the Schr\"{o}dinger equation Eq. \eqref{eq:schrodinger}.

The dispersion relation of the interface states thus reads
\begin{equation}\label{eq:dispSurf}
 E=E_{\alpha,\lambda}(k_\parallel)=\alpha \sqrt{\Tilde{E}_\lambda^2 +\hbar^2v^2k_\parallel^2},
\end{equation}
where $\alpha=\pm$ denotes the band index. This relation shows also how to convert a virtual energy to a physical energy.
%Note that $\Tilde{E}$ must be positive because otherwise $E^2 <0 $ for $k_{\parallel} = 0$. 
Since we are interested in the states localized at the surface, we consider only bound states in the quantum well defined by 
$U_{\lambda}(z)$, which, depending on the sign of the derivative $\partial_z \Delta(z)$, is confining for at least one chirality, $\lambda = -\text{sgn}(\partial_z\Delta)$. 

This is the essence of the \textit{Dirac quantum well}, which arises at a topological heterojunction: once squared, the transformed Hamiltonian $H_T$
in the Weyl basis yields two decoupled Schr\"odinger equations for an effective quantum well given by the chirality-dependent potential $U_\lambda(z)$. Moreover, the 
plane-wave motion in the $xy$-plane is decoupled from the quantized motion in the $z$-direction so that we effectively have to deal with a simple 1D quantum problem. Once solved the 1D problem, we can retrieve the spectrum of $H_T$ using Eq. \eqref{eq:dispSurf}.

\subsection{Existence of zero-energy mode and Jackiw-Rebbi argument}\label{sec:JR}

Before making an explicit mapping to the problem of a quantum well, let us just remember the Jackiw-Rebbi argument \cite{Jackiw76}, adopted by Aharonov and Casher \cite{AC79} 
in the presence of a vector potential (reminiscent of
our gap function). It states
that Eq. (\ref{eq:schrodinger0}) always hosts a chiral zero-energy solution, for $k_\parallel=0$, that is the solution of 
\begin{equation}
 \left[\Delta(z)-\lambda\hbar v \partial_z\right]\chi_\lambda^0(z)=0,
\end{equation}
which yields the massless Dirac mode with $E(k_\parallel)=\pm \hbar v k_\parallel$. This zero-energy solution
is directly obtained by integration, as long as the gap function $\Delta(z)$ is integrable,
\begin{equation}\label{eq:WFJR}
 \chi_\lambda^0(z)\sim \exp\left[\frac{\lambda}{\hbar v}\int_{z_0}^z dz'\Delta(z')\right],
\end{equation}
where $z_0$ is a reference point, which we can choose to be that where $\Delta(z_0)=0$.
One immediately sees that the solution has a definite chirality which depends on the behavior of the gap function at the 
interface. For a gap function that varies as (\ref{exp:deltaz}), only the solution with $\lambda=-$ is normalizable and represents thus the physical surface state with 
zero energy. 

To illustrate this chiral solution in a concrete example that also serves us in the discussion of the Dirac QW, let us consider the following explicit form of the gap 
function, which varies linearly over an interface of width $2l$,
    \begin{gather}
    	\Delta(z) = 
        \begin{cases}
        	-\Delta_0 & \text{if $z<-l$} \\
        	\Delta_0 \frac{z}{l} & \text{if $z \in [-l, l]$}\\
        	\Delta_0 & \text{if $z > l$},\\
        \end{cases}
        \label{exp:deltazM}
    \end{gather}
    which can for example be obtained from the linearization of a more complex behavior. Indeed, Volkov and Pankratov considered a smooth gap function 
    $\Delta(z)=\Delta_0 \tanh(z/\l)$ in their original work \cite{volkov1985two,pankratov1987supersymmetry}, but Tchoumakov \textit{et al.} showed in Ref. \onlinecite{tchoumakov2017volkov}
    that the linearized version (\ref{exp:deltazM}) of this function yields the same type of massive surface states, within the simpler framework of harmonic-oscillator function, as we 
    remind in more detail below. In the linearized case, the chiral solution (for $\lambda=-$) is given by
\begin{equation}
\label{eq:WF}
  \chi_\lambda^0\sim \left\{
  \begin{array}{ccc}
   e^{-z^2/2\xi l} & \text{for} & |z|< l\\
   e^{-|z|/\xi} & \text{for} & |z|> l, 
  \end{array}
  \right. 
\end{equation}
where 
    \begin{align}
        \xi = \frac{\hbar v}{\Delta_0}
    \end{align}
    defines an intrinsic length given in terms of the material's bulk parameters $\Delta_0$ and $v$. One thus notices a crossover from a Gaussian behavior in the interface to an exponential one outside \cite{tchoumakov2017volkov}.
    The Gaussian behavior already indicates that the solutions of the topological heterojunction with a linearized gap function are related to a parabolic confinement potential as we show in the next subsection. Notice finally that 
    a more general form of the gap function, given in terms of a corrective term $\delta\Delta(z)$, does not alter the functional form derived here, as long as this term is bounded and converges rapidly to zero outside the interface,
    i.e., for $|z|>l$.

\subsection{Witten index}\label{sec:WI}
    
    All these results are 
    in agreement with general topological arguments. It turns out that the specific form of Eq. \eqref{eq:schrodinger} stipulates the presence of a zero mode in a supersymmetric quantum mechanical 
    framework. Indeed, a 1D Schr\"{o}dinger equation such as that in Eq. \eqref{eq:schrodinger} with a potential formed by a linear combination of $\Delta(z)^2$ 
    and $\partial_z \Delta(z)$ is called the \textit{Witten equation} in the literature \cite{WITTEN1981513,pankratov1987supersymmetry} for supersymmetric quantum 
    mechanics \cite{landwehr2012landau}.
    
    To briefly develop the argument, we consider only the band extrema where $k_{\parallel}=0$. Our Bloch Hamiltonian in the Weyl basis becomes:
    \begin{equation}
    \label{eq:simplesupersym}
    H_s = - \Delta (z) \tau_x + \hbar v k_z \tau_y.
    \end{equation}
    Since $H_s$ contains only off-diagonal Pauli matrices, it maps $\chi_{+}$ to $\chi_{-}$ and vice versa when it acts on $\chi_{-}$. In the context of supersymmetric 
    quantum mechanics \cite{nakahara2018geometry}, $H_s$ plays role of the supercharge operator which relates linearly between the subspaces of fermions and bosons (here two 
    subspaces of chirality) and $\Tilde{H}=H_s^2$ is thus the supersymmetric Hamiltonian. When $\Delta(z)$ verifies Eq. \eqref{exp:deltaz}, only $\chi_-$ gets a zero-energy 
    mode with a definite chirality $\lambda=-$ while $\chi_+$ does not. We can define a quasi-topological invariant called \textit{Witten index} \cite{WITTEN1981513}:
    \begin{equation}
        \label{eq:wittenindex}
        I_{W} := \text{dim~} \text{ker} H_s |_{V_{-}} - \text{dim~} \text{ker} H_s |_{V_{+}}
    \end{equation}
    where $\text{dim~} \text{ker} H |_{V}$ is the dimension of the kernel of a linear operator $H$ acting on a subspace $V$ and $V_{\lambda}$ are the two Hilbert subspaces of 
    opposite chirality. Thus, $I_{W}$ must be an integer and invariant under continuous changes of $\Delta(z)$. It dictates also the number of zero mode, zero or one, at the 
    interface.

\subsection{Explicit discussion in terms of a Dirac quantum well}\label{sec:DQWexp}

\begin{figure}[h]
    \centering
    \includegraphics[width=0.45\textwidth]{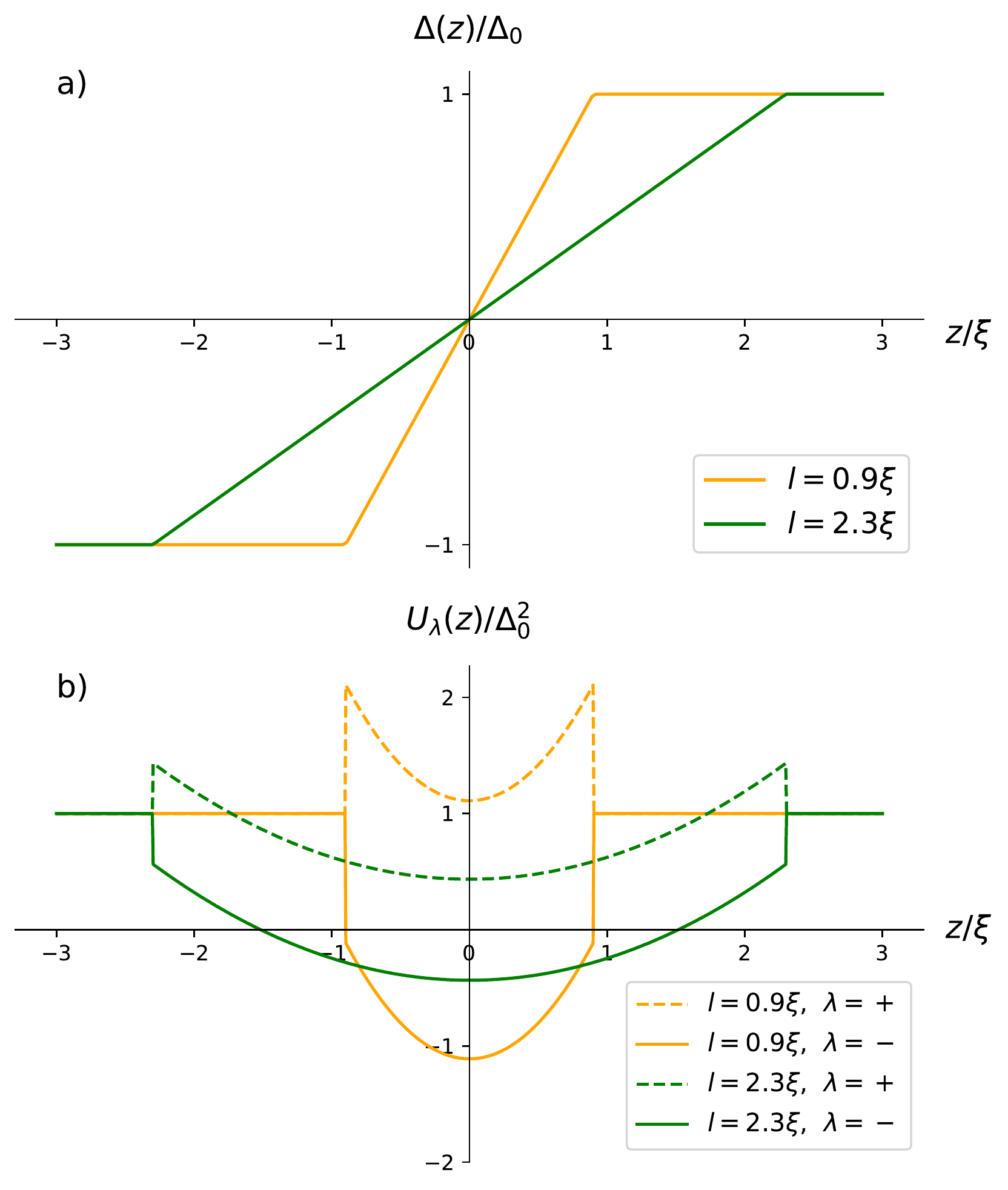}
    \caption{(a) Interface profiles described by a spatially varying gap $\Delta(z)$ for two values of characteristic interface width $l/\xi=0.9,2.3$. (b) Profiles of Dirac 
    QWs for its corresponding $\Delta(z)$ and chirality $\lambda=\pm$. $U_{-}$ is represented by solid lines and $U_{+}$ by dashed lines.}
    \label{fig:singleqw}
\end{figure}

To illustrate the Dirac QW in a concrete example, we consider again the explicit form (\ref{exp:deltazM}) of the gap function.
    Its profile is shown in Fig. \ref{fig:singleqw}(a) for two different values of the smoothness parameter $l/\xi$.
    As a result, $\Delta(z)^2$ and $\partial_z \Delta(z)$ are both even functions and $U_{\lambda}(z)$ defines thus a 
    symmetric QW potential [see Fig. \ref{fig:singleqw}(b)]. When the interface is abrupt, i.e., $l/\xi\ll 1$, one immediately sees, as already mentioned, that  
    only the fermions with $\lambda=-$ are submitted to a confining 1D QW. In contrast to this, fermions with chirality $\lambda=+$ cannot be confined in the region $z \in [-l, l]$ because they can tunnel 
    out of $z \in [-l, l]$ where the potential is no longer confining [see for example the dashed orange lines in Fig. \ref{fig:singleqw}(b)]. Thus, we can already anticipate, 
    well-known in 1D quantum mechanics, that there must be a bound state for $\lambda=-$, but not necessarily for 
    $\lambda=+$. We will show this explicitly in the following.

    %% This point can be shown generically in the framework of the Aharonov-Casher argument
    
    In the domain $z \in [-l, l]$, our Hamiltonian locally coincides with the Hamiltonian for a 1D quantum harmonic oscillator. This can be seen after formally 
    substituting $\Delta_0/v ^2\rightarrow 2m$ and $v / l\rightarrow \omega_c/2$ (or equivalently $\Delta_0/l^2\rightarrow m\omega_c^2/2$)
    in Eq. \eqref{eq:schrodinger}, such that the effective Schr\"{o}dinger Hamiltonian 
    reads
    \begin{equation}
    \label{eq:harmonic}
    \frac{\Tilde{E}_{\lambda}^2}{\Delta_0} \chi_{\lambda} = \left(-\frac{\hbar^2}{2m} \partial_z^2 + \frac{1}{2} m \omega_c ^2 z^2 + \lambda \frac{\hbar \omega_c}{2}  \right) \chi_{\lambda}.
\end{equation}
    Here, we find the Hamiltonian of a quantum harmonic oscillator with an energy shift depending on the chirality due to a vertical shift between $U_-$ and $U_+$ determined by 
    the interface width $l$. The spectrum of this Hamiltonian is thus given by \cite{tchoumakov2017volkov}
    \begin{eqnarray}
    \label{eq:specharmonic}
    \nonumber
    \frac{\Tilde{E}_{\lambda}^2}{\Delta_0} &=& \hbar \omega_c \left( n + \frac{1+\lambda}{2} \right)\\
    \quad \text{or}\qquad
    \Tilde{E}_\lambda &=& \sqrt{2\frac{\xi}{l}\left(n+\frac{1+\lambda}{2}\right)}\Delta_0
\end{eqnarray}
    where $n \ge 0$ is an integer and $\lambda = \pm$. Clearly, the Schr\"{o}dinger equation \eqref{eq:schrodinger} thus possesses exactly one zero mode for $n=0$
    in the QW confining potential $U_{-}$, in agreement with the general arguments developed in the previous subsections \ref{sec:JR} and \ref{sec:WI}
    while all other levels can in principle be accessed by both chiralities. We discuss the latter states with $n\neq 0$ in detail in the following subsection \ref{sec:MVP}.
    The zero mode, which due to the parabolic potential has exactly the Gaussian 
    shape obtained in Eq. (\ref{eq:WF}), is precisely the chiral state whose energy is immune to details 
    at the surface and independent of the well width $l$. The ground state in a single Dirac QW is thus the topological massless state at the surface of TI. Only one 
    chiral state exists because we consider only one boundary. We obtain another chiral state if we take into account the other complementary boundary.

\subsection{Massive Volkov-Pankratov states}\label{sec:MVP}
    
    In the previous subsection, we have discussed the zero-energy state $n=0$ within the explicit model of a Dirac QW.
    Besides this zero mode, Dirac QW model hosts other higher energy states for $n\neq 0$ (massive VP states) as allowed bound states for \textit{both} chiralities so that they are no longer protected from 
    back-scattering \cite{vdBerg2020}. VP states exist when the energy scale $\hbar \omega_c$ is sufficiently small compared to $\Delta_0$. Equivalently, if $l$ is much 
    larger than $\xi = \hbar v / \Delta_0$, the Dirac QW can have massive VP states besides the chiral one. The critical smoothness $l_c$ to have $n=1$ VP state is equal to 
    $\alpha \xi$ where $\alpha$ is on the order of one but its precise value depends on the precise form of $\Delta(z)$. The number of massive states which a single Dirac QW can 
    host scales as \cite{tchoumakov2017volkov} 
    
    \begin{align}
        n_{\text{max}} \approx \frac{l}{\xi}.
        \label{eq:nmax}
    \end{align}
    
    The wavefunction $\chi_\lambda$ for $n=0$ behaves as a Gaussian at the interface and decays exponentially in the bulk. The wavefunction of the bound states can 
    penetrate into the region out of the QW due to the finite well depth related to the bulk gap and the smoothness $l$. The spatial extension of the wavefunctions is thus 
    described by a length scale $l_s = \sqrt{l\xi}$ which depends on the well width and the bulk gap. As we can see from this model, the smoothness of the surface, encoded 
    in $\Delta(z) ^2$ and $\partial_z \Delta(z)$, does not only determine the width of the quantum well but it also modifies its depth [see the orange and green lines in 
    Fig. \ref{fig:singleqw}(b)]. This is essentially different from the conventional (square) QW where one can independently tune its depth and width.
    
    Note that a smoother interface gives rise to a wider but shallower Dirac QW which can nevertheless host more bound states by the thumb rule Eq. \eqref{eq:nmax}. We 
    can argue in the context of supersymmetric quantum mechanics. If a non-zero mode $\chi_{n,-}$ exist in $V_{-}$, one can find a non-zero mode of same energy in $V_{+}$ 
    by $\chi_{n-1,+}=H_s \chi_{n,-}$ because $H_s$ commutes with $\Tilde{H}$. Alternatively, if $U_{-}$ can host a bound state $n = 1$, $U_{+}$ must host a bound state $n = 0$ 
    of same energy [see Eq. \eqref{eq:specharmonic}]. Due to the tunneling effect, a bound state in $U_+$ must have a virtual energy below $\Delta_0^2$. A smoother interface with 
    larger value of $l$ reduces exactly the vertical shift between $U_{\lambda}$ and thus make the minimum of $U_+$ sink below $\Delta_0^2$ [see Fig. \ref{fig:singleqw}(b)]. 
    However, this is a necessary but not sufficient condition for the existence of bound states because the zero-point energy is finite. Yet, a smoother interface can host more bound states. 
    
    Since we consider only bound states at the interface, the argument above still applies when we consider the full global profile Eq. \eqref{exp:deltazM}. In fact, an 
    interface between a TI and a trivial one can be always modeled by Eq. \eqref{exp:deltaz} without losing much generality. For example, 
    Tchoumakov \textit{et al.} \cite{tchoumakov2017volkov} considered a profile $\Delta(z) = \Delta_0 \tanh (z/l)$ and got same conclusions. This is because the band 
    inversion mechanism allows one to linearize the spatially varying gap at the interface. Massive VP states can in principle emerge in any topological heterojunction 
    when the interface is sufficiently smooth. In other words, Dirac QW can host more bound states if it is sufficiently wide.
    
    \section{Quantum tunneling: Double Dirac quantum well} \label{sec:3}
    
\begin{figure}[h]
    \centering
    \includegraphics[width=0.45\textwidth]{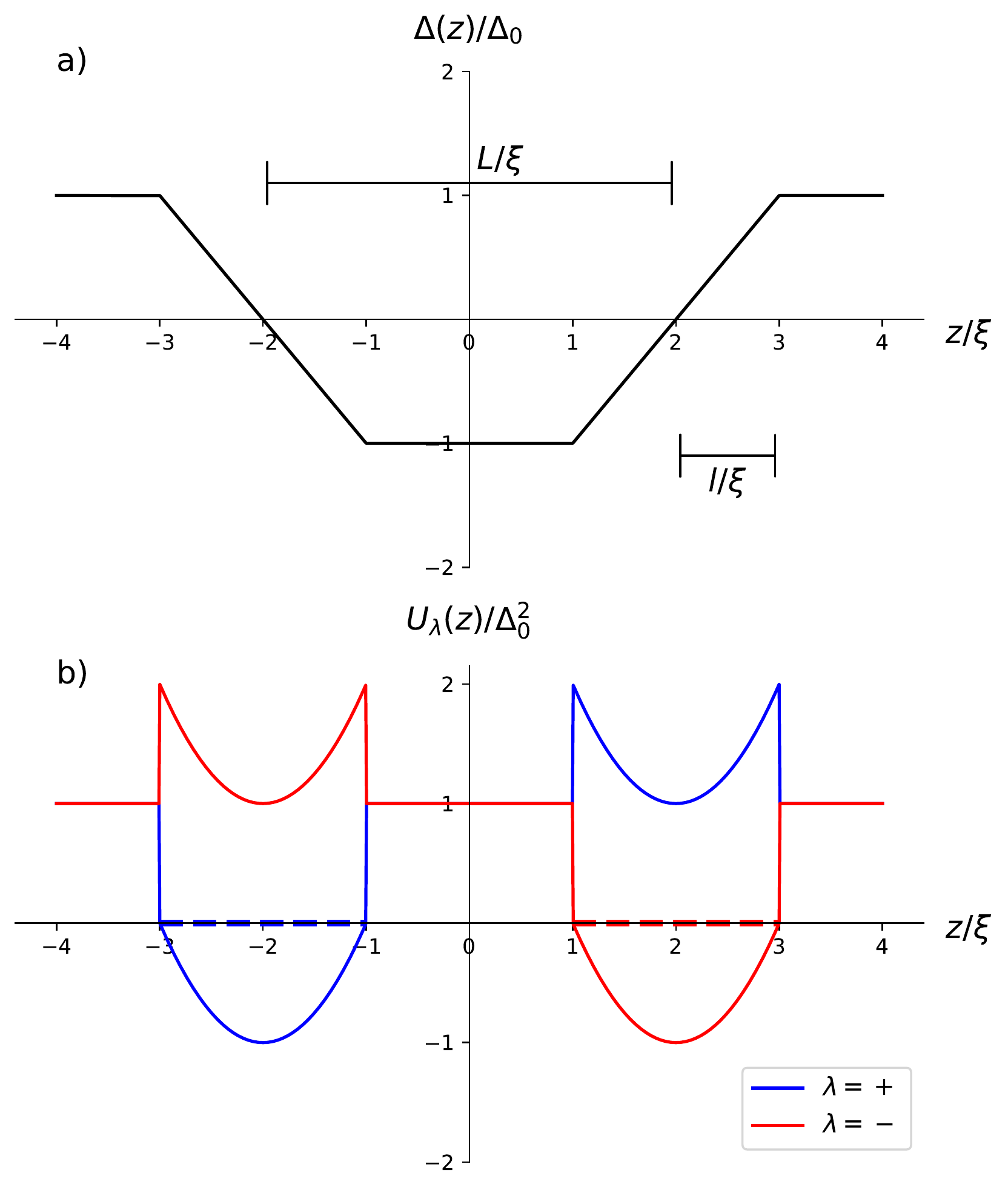}
    \caption{(a) Profile of the spatially varying gap $\Delta(z)$ for finite-size system with $l/\xi=1$ and $L/\xi=4$. (b) Profiles of two adjacent Dirac QWs for two 
    chiralities. Two dashed lines, blue and red, indicate the energy level (close to zero) of chiral states of $\lambda=\pm$, respectively.}
    \label{fig:doubleqw}
\end{figure}

We are now armed to describe the configuration of two adjacent topological heterojunctions that arises when a TI is sandwiched between two trivial insulators, as depicted
in Fig. \ref{fig:diracqw}(b). Similarly to the situation presented in the last section, this configuration corresponds, once written in terms of Eq. (\ref{eq:schrodinger}), to a \textit{double}
Dirac QW separated by a virtual energy barrier of height $\Delta_0^2$. Here, we are mainly interested in the hybridization of the surface states as a function of the width $L$ of the (thin) TI film
and the interface smoothness $l/\xi$. This hybridization can be illustrated as being due to tunneling between the two Dirac QWs, as we explicitly show in the following with the 
help of the gap function 
    \begin{gather}
    	\Delta(z) = 
        \begin{cases}
        	\Delta_0 & \text{if $z<-\frac{L}{2}-l$} \\
        	-\frac{\Delta_0}{l}(z+\frac{L}{2}) & \text{if $z \in [-\frac{L}{2}-l, -\frac{L}{2}+l]$}\\
        	-\Delta_0 & \text{if $z \in [-\frac{L}{2}+l,\frac{L}{2}-l]$}\\
        	\frac{\Delta_0}{l}(z-\frac{L}{2}) & \text{if $z \in [\frac{L}{2}-l,\frac{L}{2}+l]$}\\
            \Delta_0 &  \text{if $z>\frac{L}{2}+l$},
        \end{cases}
        \label{exp:doubledeltazM}
    \end{gather}
which enters in our model Hamiltonian \eqref{eq:h03dti} and in the effective Schr\"odinger equation \eqref{eq:schrodinger}. Notice that the width $2l$ of the Dirac QWs must naturally be 
smaller than the width $L$ of the TI film, defined as the distance between the positions where the gap function vanishes, $\Delta(z)=0$. 
Fig. \ref{fig:doubleqw} shows the form of $\Delta(z)$ along with its corresponding Schr\"{o}dinger potential $U_{\lambda}(z)$ obtained from Eq. (\ref{eq:superpot}), for $l/ \xi =1$ and $L/ \xi = 4$. 
The potential $U_{\lambda}(z)=U_{-\lambda}(-z)$ is inherently symmetric around $z=0$ upon interchange of the two chiralities and thus 
gives rise to a chiral state localized only in one of the quantum wells: $\lambda=+$ for the right QW and $\lambda=-$ for the left one. Similarly to the single Dirac QW with a gap 
function given by Eq. (\ref{exp:deltazM}), we can solve this Hamiltonian analytically, and its energy spectrum is obtained by solving numerically a {secular} equation (see Appendix). 
Notice that this equation has always plane-wave bulk solutions for energies above the bulk gap in addition to the bound states of the two Dirac QWs. This means that the hybridization of the 
surface states, which we find in exact calculations, does not only involve direct tunneling  between the bound states of two Dirac QWs, but also tunneling processes via the bulk states at energies above the gap. 

In the following parts, we discuss the spectrum of the double Dirac QW, by solving Eq. (\ref{eq:schrodinger}) then using Eq. \eqref{eq:dispSurf} for the appropriate gap function (\ref{exp:doubledeltazM}),
in comparison with that of a single one, the spectrum of which can be retrieved in the limit $L/l\rightarrow \infty$. From a tunneling point of view, the spectrum of the surface 
states is expected to be close of that for a single Dirac QW, the energies at $k_\parallel=0$ we represent henceforth by the superscript $0$, $E_n^0(k_\parallel =0)$, while the
deviation in energy is denoted by $\pm \Delta E_n$. Indeed, this deviation can be calculated with the help of the virtual energies $\Tilde{E}^2$ in Eq. (\ref{eq:schrodinger}) as
\begin{equation}
 \Delta E_n = | |\Tilde{E}| - | E_n^0(k_\parallel=0) | |,
\end{equation}
for each of the chiralities $\lambda=\pm$. In the spectrum of the surface states, this deviation has different consequences according to whether we discuss the $n=0$ state or 
the massive $n\neq 0$ VP states. Indeed, the most salient consequence is a gap opening for the $n=0$ states, which are no longer protected by the Jackiw-Rebbi argument since 
the gap function now has the same sign in both limits $z\rightarrow \pm\infty$, and the energy shift $\Delta E_0$ in the double Dirac QW model manifests itself in terms of a mass
gap in the spectrum,
\begin{equation}\label{eq:spectrum0}
 E_{\alpha,n=0}(k_\parallel)=\alpha\sqrt{\Delta E_0^2+\hbar^2 v^2k_\parallel^2}.
\end{equation}
In contrast to this situation, the massive VP states are ``split'' in energy by $\pm\Delta E_n$,
\begin{equation}\label{eq:spectrumN}
 E_{\alpha,n} (k_\parallel)=\alpha \sqrt{\left(E_n^0(k_\parallel = 0)\pm \Delta E_n\right)^2+\hbar^2v^2k_\parallel^2},
\end{equation}
as a consequence of quantum tunneling between the two Dirac QWs and the resulting hybridization of the QW states.

Two situations, a sharp and a smooth interface, respectively, are discussed in the following sections. 
All the length scales are written in units of the intrinsic length $\xi = \hbar v/ \Delta_0$.
    
\subsection{Sharp interface: $l \ll \xi$}
    \label{sec:sharp}
    
    When $l \ll \xi$, only the chiral states with $n=0$ are obtained as localized solutions at two spatially separated interfaces. They cannot hybridize directly due to their opposite chiralities that 
    give rise to a vanishing scalar product between their respective spinorial wavefunctions. However, we obtain in our exact calculations a mass-gap opening in the spectrum
    since even the $n=0$ QW states are no longer forced by the Jackiw-Rebbi argument to remain at zero energy. Indeed, the gap function changes its sign twice such that the 
    gap has the same sign on the far left hand side of the double QW as well as on the far right hand side. A function of the form (\ref{eq:WFJR}) would thus no longer be normalizable. In the double QW model [see Eq. \eqref{eq:schrodinger}] this
    can be understood in perturbation theory. Indeed, the wavefunctions $\chi_{\lambda, n=0}(z)$ are sensitive, via their exponential tail 
    to the modified well potential even if it is situated at higher energies (above the virtual energy $\Delta_0^2$). Within our exact calculations, we can obtain the deviation
    in energy and thus the mass gap, in the limit $l\ll \xi$ of a sharp interface, by an expansion of the secular equation Eq. \eqref{eq:secular} in terms of  $l/\xi$ 
    (see Appendix). This yields 
\begin{equation}
    \label{eq:splittingchiral}
    \Delta E_{0} = \Delta_0 e^{-\frac{L}{\xi}} \sqrt{ 1+ \frac{4l^2}{3\xi^2} } .
\end{equation}
The exponential decay with increasing thickness $L$ in our formula agrees with previous theoretical results on thin films of TIs with sharp 
surfaces \cite{lu2010massive,liu2010oscillatory,linder2009anomalous,Shan_2010}. When $L> \xi$, the gap in the surface spectrum is an exponentially decreasing function of the bulk gap parameter $\Delta_0$,
$\exp(-L/\xi)=\exp(-L\Delta_0/\hbar v)$, such that one can say that the chiral states is protected by the bulk gap. Furthermore, we obtain another algebraic correction in $(l/\xi)^2$
that stems from the smoothness of the interface. For a Bi$_2$Se$_3$ 
thin film of four quintuple layers (QLs), Neupane \textit{et al.} have measured an energy gap of the surface Dirac cone 0.05 eV for a sample of 4 
QLs \cite{neupane2014observation}. Remarkably, we find a very close value of $2 \Delta E_0 = 0.03$ eV by our recipe even though $l$ equals 
at least the thickness of a QL ($\sim 1$ nm), which is of the same order as the 
intrinsic length $\xi$. For Bi$_2$Se$_3$, $\xi$ is 1.5 nm taking $2\Delta_0 = 0.35 \  \eV$ and $v = 2.5 \  \eV \cdot $\AA \cite{bianchi2010coexistence,xia2009observation}.
The discrepancy between our calculated value and the experimentally measured one could be attributed to the particle-hole symmetry-breaking term $k^2$ which we do not 
include in our model. Another possible origin of the slight mismatch between the values could be an asymmetry in the experimental quantum wells where the interface 
thickness is not always the same, i.e., $l_-\neq l_+$ at $\mp L/2$ in our approach where $l_-$ for the left and $l_+$ for the right interface.
We emphasize that the main advantage of our formula is that one can estimate the chiral state splitting with rather reliable precision with simple analytical calculations. 
One can also use our formula to deduce the characteristic length $l$ from the energy splitting of the surface Dirac cone. We estimate $l \sim 2$ nm for Neupane \textit{et al.}'s 
sample.

    \subsection{Smooth interface: $l > \xi$} 
    \label{sec:smooth}
    
    Let us now consider the more interesting situation of smooth interfaces when $l>\xi$ and massive VP states emerge in addition to the chiral state. The magnitude of the energy 
    splitting 
    depends on $l/\xi$, $L/\xi$ and the VP state index $n$ for a given set of Fermi velocity and bulk gap. In the following, we first fix $L/\xi = 20$ and change $l/\xi$. 

\begin{figure}[h]
    \centering
    \subfigure{
    \includegraphics[width= 0.45\textwidth]{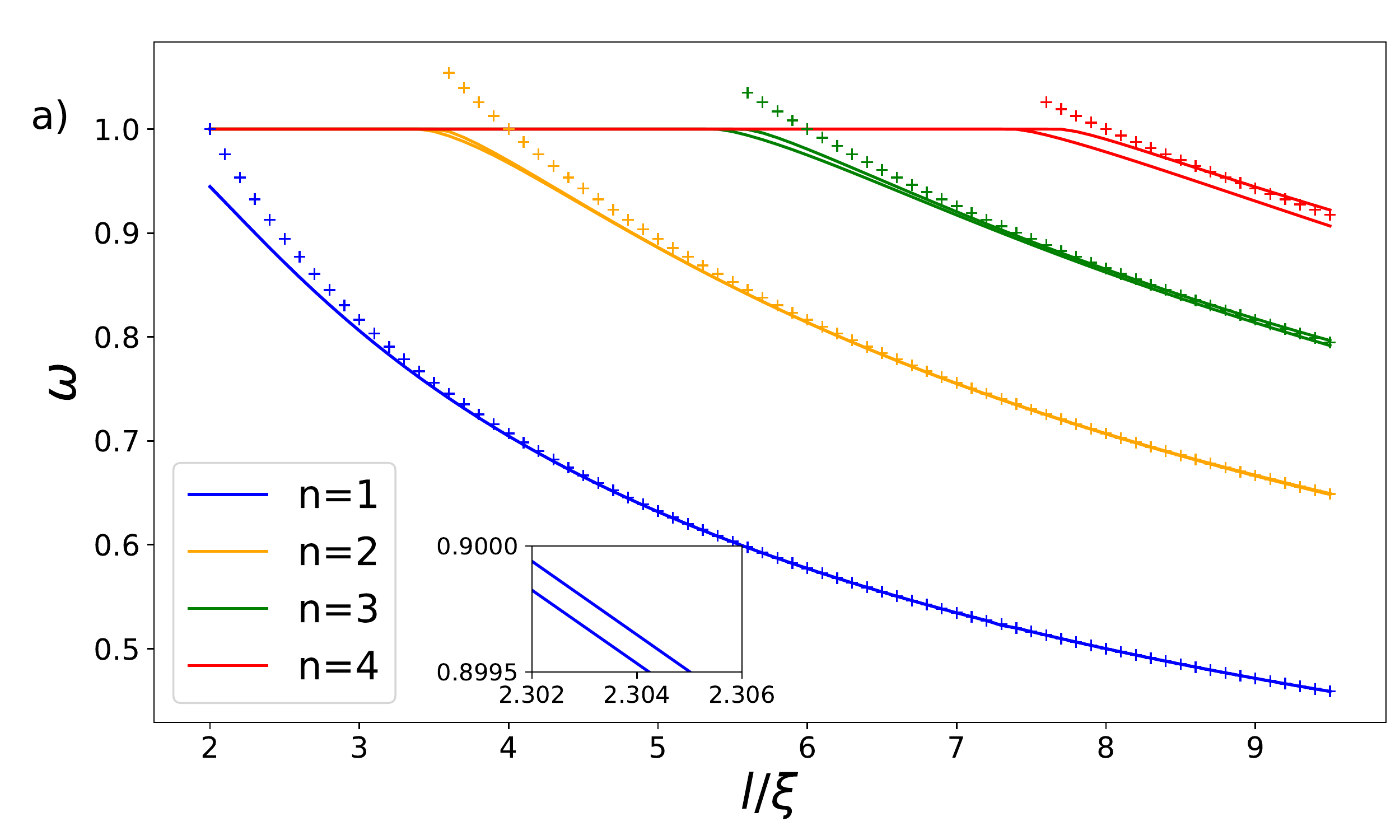}
    }
    \subfigure{
    \includegraphics[width= 0.45\textwidth]{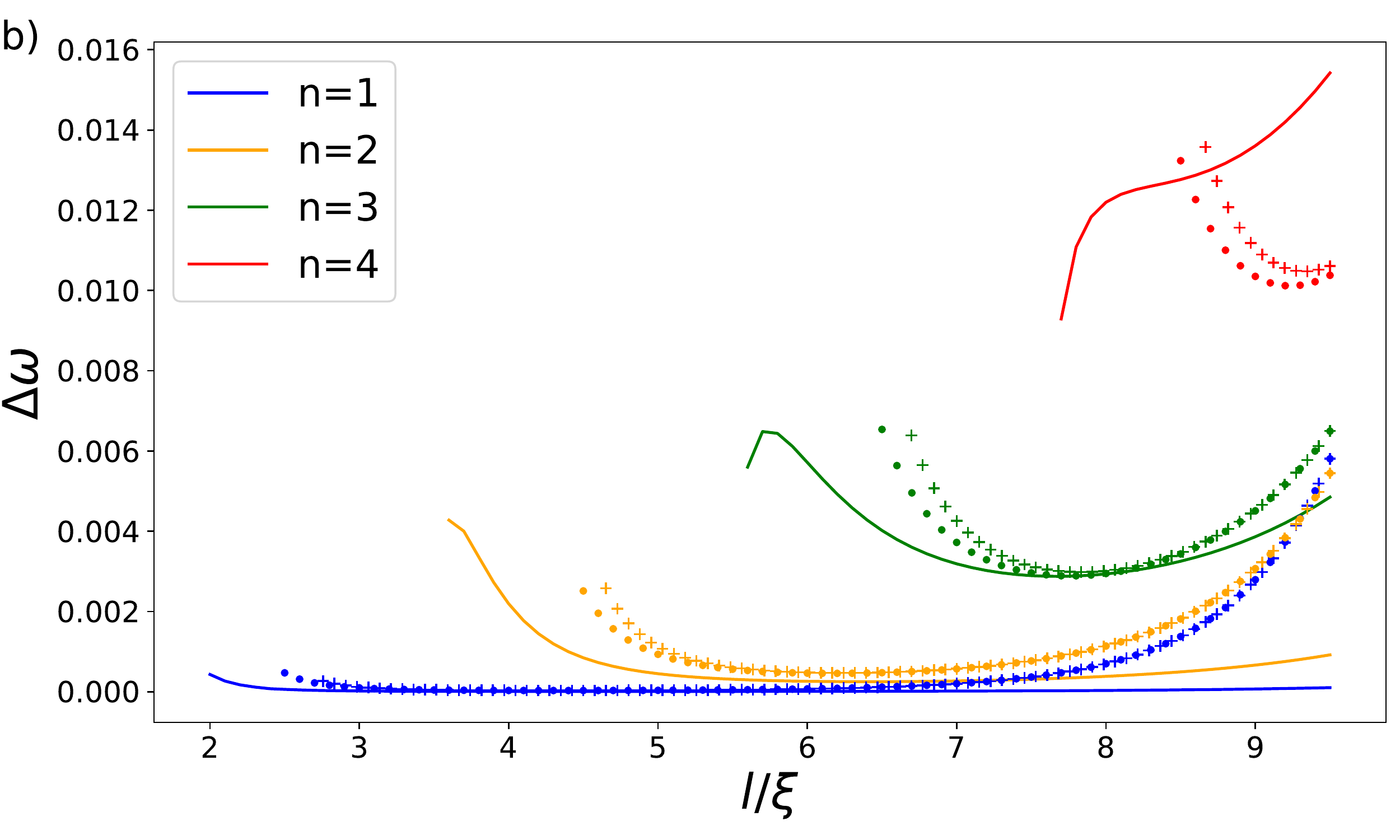}
    }
    \caption{(a) Reduced energy $\omega$ as a function of smoothness $l$ for the massive VP states $n=1$ to 4 by solving the secular equation Eq. \eqref{eq:secular} (solid lines). The $l$-dependence of $\omega$ is well described by Eq. \eqref{eq:linearapprox} (crosses). Inset: zoom-in to show the splitting for the VP state $n=1$. 
    (b) Reduced energy splitting $\Delta \omega$ as a function 
    of smoothness $l$ for the massive VP states $n=1$ to 4. The distance between two Dirac QWs is set to be $L/\xi=20$. The continuous lines show the splitting obtained from
    our secular equation Eq. (\ref{eq:secular})
    of the double Dirac QW problem, and the crosses indicate the values obtained from the approximate formula (\ref{eq:splittingsquareB_vp}) based
    on quantum tunneling between the QWs. The dotted lines show results based on the same formula, where we have used the exact energies 
    for the VP states of a single Dirac QW instead of the approximate ones given in Eq. (\ref{eq:linearapprox}).
   }
    \label{fig:width}
\end{figure}

Fig. \ref{fig:width}(a) shows the results of our calculation, based on Eq. (\ref{eq:secular}),
for the variation of $\omega$ as a function of $l/\xi$ for the massive VP until $n=4$. We define 
here $\omega$, a reduced QW energy
\begin{align}
    \centering
    \label{exp:omega}
    \omega = \sqrt{\frac{E^2 - \hbar^2 v^2 k_{\parallel}^2}{\Delta_0^2}} \in [0,1]
\end{align}
which is the energy at $k_{\parallel}=0$ in units of half of the bulk gap $\Delta_0$. %{\color{red} *** est-ce qu'on a besoin d'introduire $\omega$ ? ***}
For the given set of parameters, the gap opening of the chiral state is $10^4$ 
times smaller than the energy splitting of the massive VP states, which is another manifestation of topological protection (see discussion below). 
We first notice in Fig. \ref{fig:width}(a) that the massive VP states still follow to great accuracy the behavior
\begin{equation}
\label{eq:linearapprox}
 \omega \simeq \sqrt{2n \frac{\xi}{l}}
\end{equation}
expected from Eq. (\ref{eq:specharmonic}) for a linearized gap function in the case of a single topological heterojunction [see crosses in Fig. \ref{fig:width}(a)]. As expected, the approximation becomes less accurate 
at energies close to the bulk gap $\Delta_0$, where one notices a deviation and, most prominently, a splitting of the energies.

\begin{figure}[h]
    \centering
    
    \includegraphics[width= 0.45\textwidth]{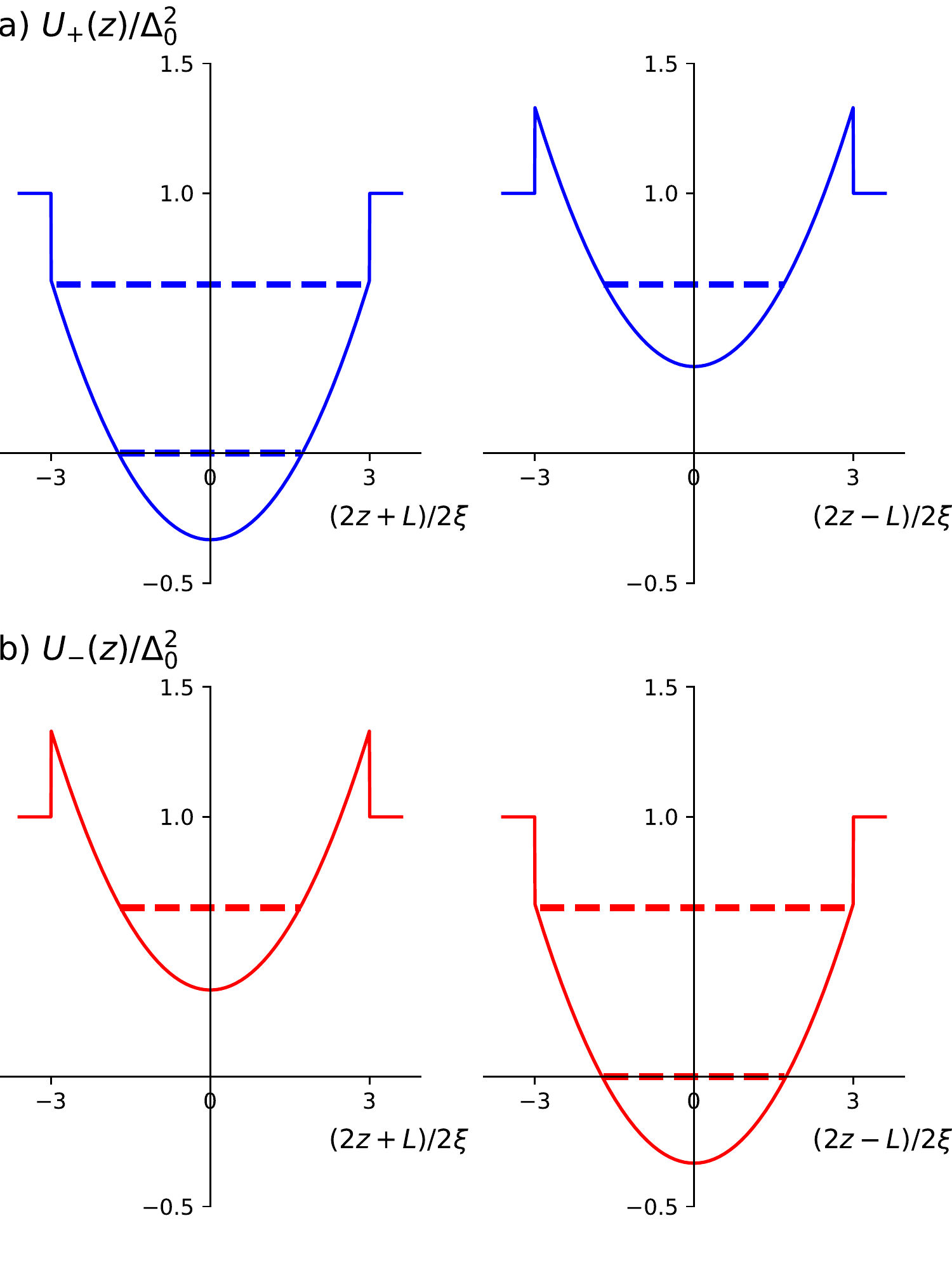}
    
    \caption{Profiles of two infinitely separated Dirac QWs for two 
    chiralities with $l/\xi = 3$: (a) for $\lambda=+$ and (b) for $\lambda=-$. The dashed lines, blue and red, indicate the energy level of chiral states and $n=1$ VP states for $\lambda=\pm$, respectively.}
    \label{fig:doubleqw_l3}
\end{figure}

Our results can be understood 
in the framework of the asymmetric double quantum well for a given chirality, as shown in Fig. \ref{fig:doubleqw_l3}, where two Dirac QWs of $\lambda=\pm$ with $l/\xi = 3$ are far 
away from each other. Thus, the tunneling effect is negligible and two VP states, each of which is situated in one of the QWs, respectively, are degenerate. Imagine now that we bring two Dirac QWs together progressively and the tunneling strength increases to lift the degeneracy of the VP states. Since the tunneling effect between two states is strongest 
when they have the same energy, the zero mode in one Dirac QW is protected because its adjacent Dirac QW does not have a zero mode of the same chirality and other VP 
states are far from the chiral state in energy. In other words, two zero modes with $n=0$ are present at each of the two interfaces, but they are protected from tunneling-induced hybridization thanks to its well-defined chirality.

%. One can only invoke the coupling to the higher energy states such as the bulk to indirectly introduce a coupling between two zero modes, as we have already mentioned in the context of the sharp interface. 
In contrast to the $n=0$ states, the massive VP 
states can hybridize strongly because they have their partner of the same energy and chirality in the adjacent QW. This is shown in Fig. \ref{fig:width}(b), where we represent the energy 
splitting of the $n\neq 0$ VP states by solving the secular equation in our double Dirac QW model (solid lines). In order to understand these results in the light of 
tunneling events between the QWs, we can heuristically derive a formula of the energy splitting due to the tunneling between adjacent finite symmetric square QWs \cite{basdevant}
\begin{equation}
    \label{eq:splittingsquare}
    2\Delta E = \frac{\hbar^2 \pi^2}{4 m l^2} \frac{4 e^{-K(L-2l)}}{2 K l},
\end{equation}
where $2l$ is the width of a square QW, $L$ the separation between the centers of the
two square QWs and $K = \sqrt{2mV_0}/\hbar$ with the effective depth of the square QW, $V_0$. As we did above in the derivation of Eq. \eqref{eq:harmonic}, 
we replace $2m$ by $1/v^2$ and $V_0$ by $(1-\omega^2)\Delta_0^2$ for the effective potential depth for a surface state of reduced energy $\omega$ given by Eq. (\ref{exp:omega}). 
Notice that, in our Dirac QW approach, the energy splitting is that of the virtual energies. In order to translated this splitting into one of the physical energies,
one needs to take into account an expansion of the virtual energies to linear order around the VP energies $\omega$, here.
The wave vector that describes the exponential suppression of the hybridization is given by 
\begin{equation}\label{eq:wavevector}
 K=\frac{1}{\xi}\sqrt{1-\omega^2}.
\end{equation}
In the case of wide Dirac QWs or smooth interfaces, we can use the linearized version for the energy of the VP states (\ref{eq:linearapprox})
so that our heuristic formula reads 
\begin{equation}
\label{eq:splittingsquareB_vp}
\Delta E_n = \frac{\pi^2}{4} \frac{\Delta_0}{\sqrt{2n}} \left(\frac{\xi}{l}\right)^{5/2}  \frac{ e^{-\sqrt{1-\frac{2n \xi}{l}}(L-2l) / \xi} }{\sqrt{1-\frac{2n \xi}{l}}}
\end{equation}
for the VP states.

In Fig. \ref{fig:width}(b), we compare the energy splitting of the VP states obtained from the secular equation (solid lines) and by the heuristic formula 
Eq. \eqref{eq:splittingsquareB_vp} ($+$ symbols). The heuristic formula gives a good order of magnitude for the splitting, especially for $n=1$ and $2$. Most saliently, 
the non-monotonic behavior of the splitting as a function of $l$ is also captured by Eq. \eqref{eq:splittingsquareB_vp}. The reason is that the energy of the massive VP states 
decreases when $l$ becomes larger [see Eq. \eqref{eq:linearapprox}] as for a usual quantum well when the well width increases. However, larger values of $l$ shorten 
the effective separation between two QWs, $L_{\text{eff}} = L -2l$. These two effects compete with one another and give rise to a minimal value of the energy splitting when we bring the two QWs together. 
However, Eq. \eqref{eq:splittingsquare} is only valid for the bound states at the bottom 
of the square QW. So our heuristic formula is also only valid for the VP states with small $n$ [see $n=1,2,3$ in Fig. \ref{fig:width}(b)] and large $l/\xi$. 
Otherwise, even the approximation \eqref{eq:linearapprox} is no longer valid. But using the exact energies does not improve much the results [see dotted lines in Fig. \ref{fig:width}(b)]. In the case of small values of both
$n$ and $l/\xi$, the results by the heuristic formula are 
exceedingly wrong because the virtual energy level is close to the virtual energy edge $\Delta_0^2$ of the Dirac QW, i.e., the VP states just emerge from the bulk gap. 
For the same reason, higher VP states can have a monotonic behavior with increasing $l$ 
[see $n=4$ in Fig. \ref{fig:width}(b)]. Another possible origin of the quantitative discrepancy between our results obtained the secular equation Eq. \eqref{eq:secular} and those given by 
Eq. (\ref{eq:splittingsquareB_vp}) stems from the form of the wavefunctions inside each quantum well. While our model shows that these wavefunctions are given by the
harmonic-oscillator functions (a Gaussian combined with a Hermite polynomial), Eq. (\ref{eq:splittingsquareB_vp}) has been obtained for a square well potential in which case the 
wavefunctions inside are sine and cosine functions. 

Compared to the problem of one Dirac QW, the critical values of $l_n$ in the double Dirac QW, above which the $n-$th massive VP states appear in the gap, are almost the same 
as those of the single Dirac QW. For example, for $L/\xi=20$, the critical value of $l_n/\xi$ for $n=2$ is around $3.6$ while $l_n/\xi = 3.7$ in the case of single Dirac QW. 
This means that the smoothness of the surface is a local property of the surface. However, due to the splitting, the first appearance of a massive VP state requires a slightly 
smaller value of $l$ for the double Dirac QW than for the single Dirac QW. This can lead to a situation where the lower energy state of the $n-$th VP states exists in the 
gap but the higher one does not.

Let us now fix $l=6\xi$ varying $L/\xi$ to study how the splitting of VP states depends on the thickness $L$. Notice that the splitting of the $n=0$ states is again negligible here.
\begin{figure}[h]
    \centering
    \includegraphics[width=0.45\textwidth]{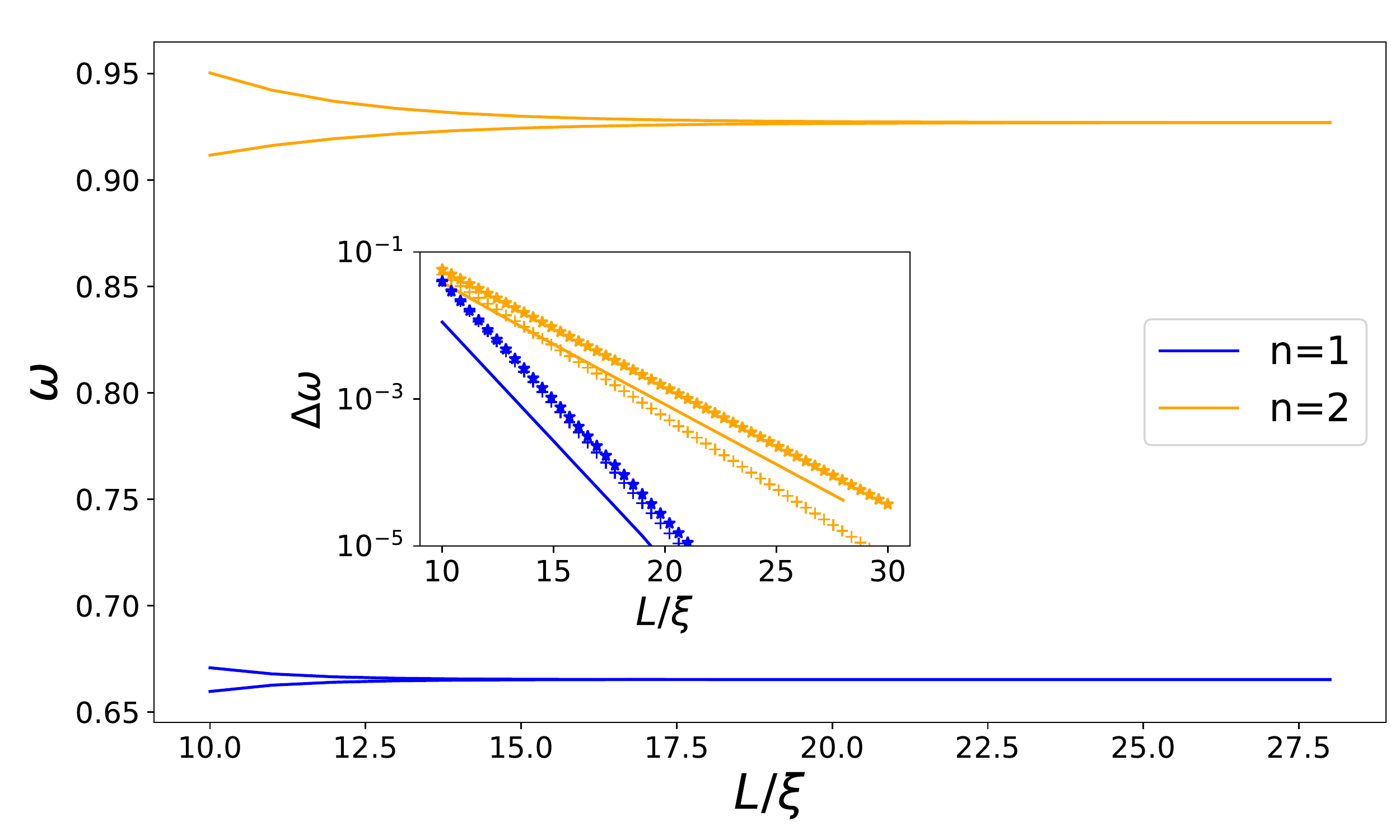}
    \caption{Reduced energy $\omega$ as a function of the distance between two Dirac QWs, $L$, for the massive VP states $n=1$ and 2. The results are obtained by solving the secular equation Eq. (\ref{eq:secular}). 
    Inset: the reduced energy splitting $\Delta \omega$ 
    decays exponentially with increasing $L$, only shown for the massive VP states $n=1$ and 2. The continuous lines indicate the splitting from the secular equation, while 
    the crosses represent the results based on Eq. (\ref{eq:splittingsquareB_vp}). The dotted lines show results based on the same formula, where we have used the exact energies 
    for the VP states of a single Dirac QW instead of the approximate ones given in Eq. (\ref{eq:linearapprox}).}
    \label{fig:thickness}
\end{figure}
Fig. \ref{fig:thickness} shows the variation of $\omega$ changing $L/\xi$ for VP states $n=1$ and 2. As in the conventional double square QW, the energy splitting due to the 
tunneling effect is exponentially weak when we increase the distance between two QWs. This is shown in the inset of Fig. \ref{fig:thickness}, where we compare our results from the secular equation Eq. \eqref{eq:secular} (continuous lines) with that obtained from Eq. (\ref{eq:splittingsquareB_vp}) (crosses). One notices that the latter approximate formula provides the
correct order of magnitude of the splitting, but overestimates it for $n=1$. Some reasons for it have already been invoked above: first, the linear-gap-function approximation 
provides energies that are less reliable when we approach the bulk-gap edge; and second, the heuristic formula (\ref{eq:splittingsquareB_vp}) has been obtained for 
a square quantum well, while the potential used in our calculations is parabolic [see Fig. \ref{fig:doubleqw_l3}]. It is likely the latter that is at the origin 
of the quantitative discrepancy. Indeed, we have compared our results to an approximate equation for the splitting, where we have used in Eq. (\ref{eq:splittingsquareB_vp}) the 
energies $\Tilde{E}^0_\lambda$ of the \textit{single} Dirac QW instead of the approximate energies given by Eq. (\ref{eq:linearapprox}). The results for the splitting obtained under
this assumption are shown in the inset of Fig. \ref{fig:thickness} in the form of the dotted lines. While this approximation better fits with the slope of the splittings obtained from the secular equation \eqref{eq:secular}, it continues to systematically overestimate the quantitative value of the energy splitting.

\begin{figure}[h]
    \centering
    \subfigure{\includegraphics[width=0.45\textwidth]{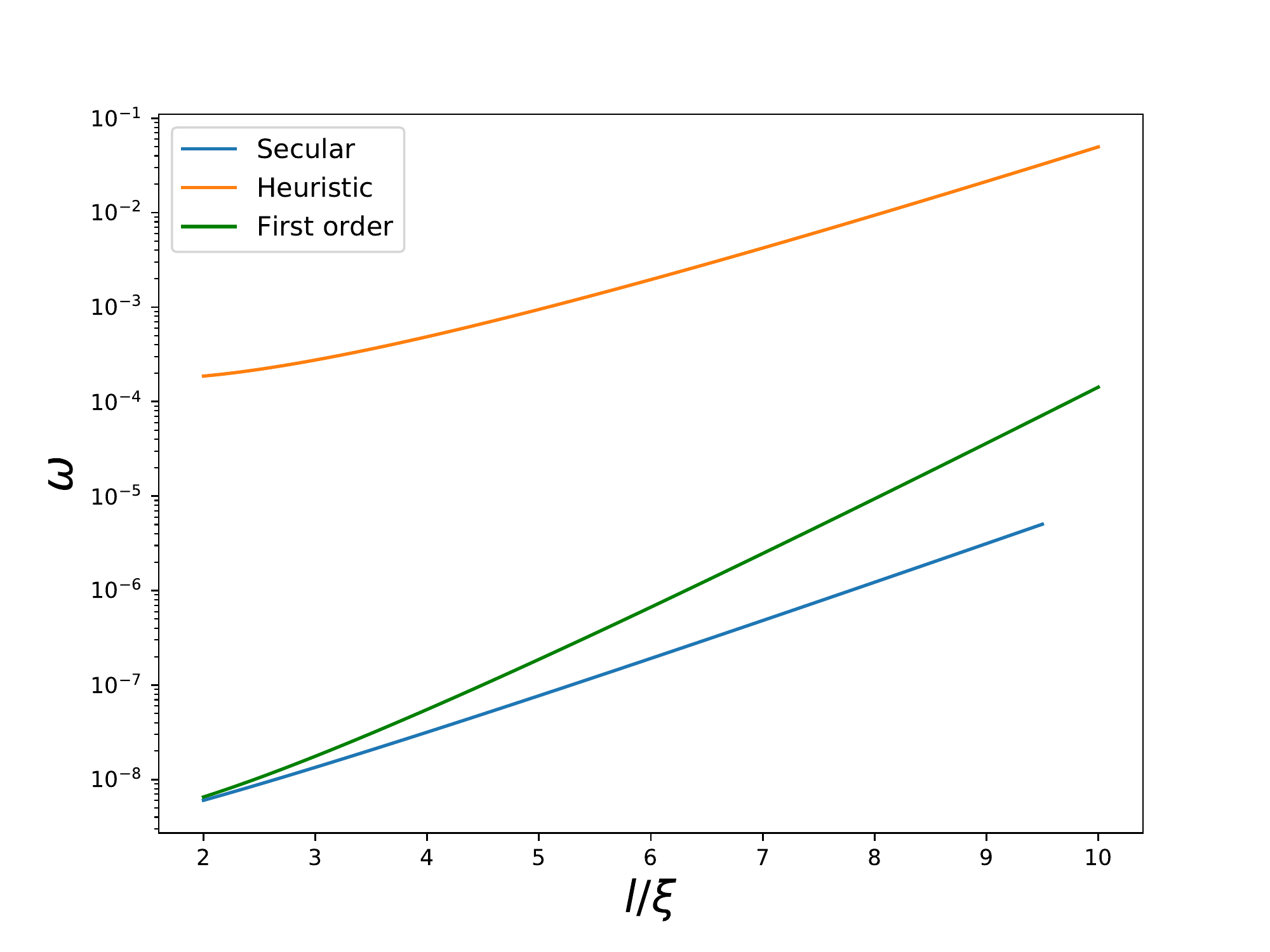}}
    \subfigure{
    \includegraphics[width= 0.45 \textwidth]{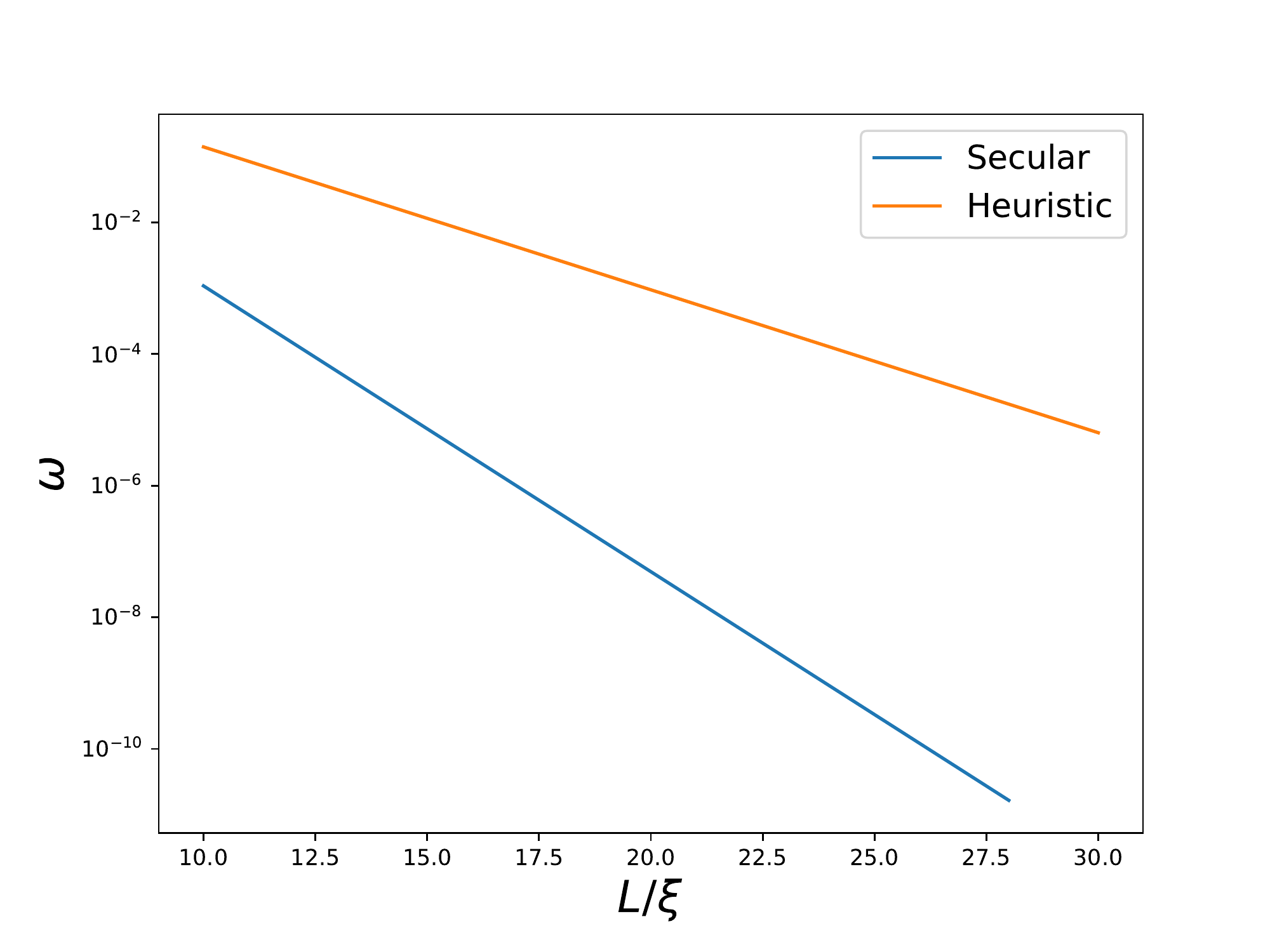}
    }
    \caption{Top: $\log_{10} \omega$ vs. $l$ for the $n=0$ chiral state and $L/\xi=20$. Bottom: $\log_{10} \omega$ vs. $L$ for the $n=0$ chiral state and $l/\xi=6$. 
    The results from the secular equation Eq. \eqref{eq:secular} are represented by blue lines, those using Eq. \eqref{eq:splittingsquareB_chiral} 
    by orange lines and those using Eq. \eqref{eq:first_order} by green line.}
    \label{fig:widthchiral}
\end{figure}

We finish this section with a short discussion of the chiral $n=0$ states. As mentioned in Sec. \ref{sec:sharp}, the energy shift yields a mass gap of the chiral states given by the energy 
$2 \Delta E_0=2 \omega \Delta_0$, the logarithm of which is plotted as the blue lines in Fig. \ref{fig:widthchiral}, based on the solution of the secular equation Eq. \eqref{eq:secular} for the double Dirac QW problem. 
From the slope of the curves, we can infer that the mass gap scales approximately as 
\begin{equation}
 E_0 \sim e^{-(L-l)/\xi},
\end{equation}
i.e., the effective width of the potential barrier is now $L_\text{eff}^0=L-l$ instead of $L_\text{eff}=L-2l$, as for the massive VP states. In order to corroborate the different
underlying physical properties of the $n=0$ state as compared to the VP states, let us consider first of all -- erroneously as we show below -- a similar 
tunneling formula as Eq. (\ref{eq:splittingsquare}), in which case we would
obtain 
\begin{equation}
\label{eq:splittingsquareB_chiral}
    \Delta E_0 = \frac{\pi}{2} \Delta_0 \left(\frac{\xi}{l}\right)^{3/2} e^{-(L-2l) / 2\xi}
\end{equation}
for the chiral state, upon the same substitutions as for the massive VP states. Notice that, in contrast to the tunneling formula for the massive VP states, there
is no linearization of the virtual energies involved here because the originally chiral surface states with $n=0$ are situated at zero energy. The splitting in energy is therefore 
obtained directly by taking the square root of the virtual energies in our Dirac QW model. This is reflected by the factor $1/2$ 
in the exponent of Eq. (\ref{eq:splittingsquareB_chiral}) as compared to the analogous expression (\ref{eq:splittingsquareB_vp}).
That Eq. (\ref{eq:splittingsquareB_chiral}) is now based on shaky grounds should be clear from the fact that the $n=0$ state 
has no partner of the same chirality in the ``other'' QW, but one would need to invoke a coupling in the form of a perturbative term that relates the two chiralities $\lambda$. This 
provides a stronger protection of the $n=0$ states, a \textit{topological} protection, than that of the massive VP states. Indeed, we have plotted the splitting expected on 
the basis of Eq. (\ref{eq:splittingsquareB_chiral}) with the orange lines. One immediately notices that a tunneling-induced splitting mechanism overestimates the correct mass gap by several orders of magnitude in the full range of values $L/\xi$ and $l/\xi$ that we have investigated. 

However, we can get a better agreement in perturbation theory. The leading order of perturbation to open a gap is given by the process that the chiral state is weakly affected by the deviation of the potential $U_\lambda(z)$
from $\Delta_0^2$ in the exponential tail. To illustrate this point, let us consider the chirality $\lambda=+$, in which the $n=0$ state is located in the left QW, in terms of a 
wavefunction (\ref{eq:WF}) but now centered around $z=-L/2$ [see Fig. \ref{fig:doubleqw_l3}(a)]. This wavefunction represents the exact zero-energy state when the QW potential is constant when $z > -L/2 + l$ so that $U_+(z>-L/2+l)=\Delta_0^2$, i.e., when there is no second QW. The other QW at $z=L/2$ therefore gives rise to a deviation 
\begin{equation}
\Delta U_+(z)=\Delta_0^2\left[\left(\frac{\xi}{l}-1\right)+\left(\frac{2z-L}{2l}\right)^2\right],
\end{equation}
and the deviation in energy of the zero mode can be calculated as 
\begin{equation}
\label{eq:first_order}
 \Delta E_{0}^2=\int_{-\infty}^{\infty}dz\, \chi_+^{0*}(z) \Delta U_+(z) \chi_+^0(z).
\end{equation}

In terms of $\Delta \omega^2$, the formula reads
\begin{equation}
\label{eq:first_order_no_approx}
 \Delta \omega^2= A^2 \frac{\xi^3}{2l^2} e^{- \frac{2L-l}{\xi}} \left[ \sinh \left(\frac{2l}{\xi} \right) - \frac{2l}{\xi} e^{-\frac{2l}{\xi}} \right]
\end{equation}
where $A$ is the normalization factor of the wavefunction $\chi_+^{0}$:
\begin{equation}
    \label{eq:normalization_factor}
    A^{-2} = \sqrt{\pi l \xi} \   \text{erf} \left(\sqrt{\frac{l}{\xi}} \right) + \xi \ e^{-\frac{l}{\xi}}
\end{equation}
where erf$(x)$ is the error function. When $l/\xi \gg 1$, we have
\begin{equation}
    \label{eq:large_l_chiralsplit}
    \Delta E_0 = \frac{\Delta_0}{2 \pi^{1/4}} \left(\frac{\xi}{l}\right)^{5/4} e^{- \frac{L_{\text{eff}}}{\xi}}
\end{equation}
where $L_{\text{eff}}$ is now $L - 1.5 l$. We easily remark that this formula captures the exponential decay of $E_0$ as a function of $L/\xi$. In the other hand, the formula (\ref{eq:large_l_chiralsplit}), though limited at the first order of perturbation, gives a rather good approximation to the result by the secular equation Eq. \eqref{eq:secular} especially when $l/\xi$ is not too large (green line in Fig. \ref{fig:widthchiral}). The reason for the discrepancy is that higher order contributions in perturbation theory are non-negligible when $L_{\text{eff}}$ becomes smaller. In a tunneling point of view, since the energy spacing between the $n=1$ VP states and the chiral state is a decreasing function of $l/\xi$ [see Eq. \eqref{eq:linearapprox}], the hybridization between them is thus stronger with increasing $l/\xi$.

Notice that we have discussed, here, only hybridization between states of the same index $n$. This is indeed the most interesting situation in symmetric QWs, where the 
left and right interfaces have the same width such that the surface states associated with the two different interfaces have the same energy. In this resonant situation, 
hybridization is strongest, while in the case of different interface widths the degeneracy of the VP states is trivially lifted. As one can see from Eq. (\ref{eq:specharmonic}), 
another resonant situation may arise between VP states of different index 
($n_-$ for the left and $n_+$ for the right interface) for well chosen ratios $l_-/l_+\simeq n_-/n_+$ between the width parameters of the left and right interfaces, respectively. 
However, a detailed discussion of this situation is beyond the scope of the present paper.

\section{Conclusions}

In this paper, we have investigated the surface states of a thin TI sandwiched between two trivial insulators as a function of interface smoothness and the width of the TI slab. 
This situation is conveniently described in terms of two decoupled Schr\"odinger equations, one for each chirality, of a 1D quantum particle in a double QW. As we have shown
along the lines of Ref. \cite{tchoumakov2017volkov},
the QW structure, which we call \textit{Dirac QW} here, arises when one squares the Hamiltonian describing a Dirac particle that changes its mass gap from positive to negative 
values in a topological heterojunction. If the sign change is smooth in the interface, over a characteristic length $l$ that is larger than the intrinsic length $\xi=\hbar v/\Delta_0$
in terms of the bulk parameters $v$ (velocity) and $\Delta_0$ (half of the bulk gap), one finds massive VP states ($n\neq 0$) in addition to the usual chiral one ($n=0$) \cite{volkov1985two,pankratov1987supersymmetry,tchoumakov2017volkov}. Our main interest resides in the fate of these states once we consider both interfaces, i.e., two coupled 
topological heterojunctions that give rise, upon squaring of the Hamiltonian, to a \textit{double} Dirac QW problem with an asymmetric well potential that respects 
$U_\lambda(z)=U_{-\lambda}(-z)$ upon interchange of the chiralities $\lambda$. The massive VP states behave extremely differently as compared to the topological chiral states. Indeed, 
each massive VP state in one of the wells has a partner of the same chirality $\lambda$ in the other one. Since we have considered a symmetric situation, where both interfaces
have the same smoothness $l$, these partners have the same energy and are separated by an energy barrier of an effective width $L_\text{eff}=L-2l$. The energy splitting 
of these VP states, which is obtained by solving the secular equation for the double Dirac QW, can to great extent be understood as induced by quantum tunneling between these states in 
the two quantum wells, with an exponential behavior $\Delta E\sim \exp[-(L-2l)/\xi]$. Possible quantitative discrepancies have been identified as due to the form of the QW potentials.

Notice that, apart from transport measurements in HgTe/CdHgTe heterostructures \cite{inhofer2017observation}, there are no direct spectroscopic observations of massive
VP states at the surfaces or interfaces of topological materials. Additional surface states have, however, been found in Bi$_2$Se$_3$ \cite{bianchi2011prl} and Bi$_2$Te$_3$ 
\cite{chenchaoyu2012pnas} when the materials are exposed to an oxidizing atmosphere that renders the surfaces effectively smooth, and the orders of magnitude indicate that an 
interpretation in terms of VP states is reasonable. A clear-cut experimental verification of our theoretical findings would require heterostructures between trivial and topological 
insulators, in which the chemical composition, e.g. in a growth process by molecular beam epitaxy, varies smoothly over a certain distance in the nm range.

The fate of the chiral states is strikingly different from that of the massive VP states. While they are no longer constrained at zero energy, since the 
Jackiw-Rebbi argument no longer applies here, the 
induced mass gap in these states is \textit{not} due to quantum tunneling since the $n=0$ in one QW does not have a partner of the same chirality $\lambda$ in the other QW. 
Quantum-tunneling induced mass gaps or splittings would therefore require a coupling between the different chiral sectors. A similar formula for quantum tunneling, as that for $n\neq 0$ massive VP states, overestimates the mass gap of the $n=0$ states by several orders of magnitude. The heuristic formula Eq. \eqref{eq:splittingsquareB_chiral} cannot describe this behavior of $n=0$ chiral states as well as Eq. \eqref{eq:splittingsquareB_vp} does for massive VP states.
Furthermore, we find that the mass gap scales as $\Delta E_0 \sim \exp[-(L-l)/\xi]$, i.e., with an effective barrier width of $L_\text{eff}^0=L-l$ rather than $L-2l$. Although the formula Eq. \eqref{eq:large_l_chiralsplit} by first-order perturbation theory gives a not too bad estimation for $\Delta E_0$, its effective barrier width is $L_{\text{eff}}= L - 1.5 l$ instead of $L-l$.
The necessity of a coupling between the different chiral sectors for a substantial mass-gap opening of the $n=0$ states in the Dirac QW model is a complementary understanding of the usually invoked \textit{topological protection} of these states.

\begin{acknowledgements}
We acknowledge financial support from Agence Nationale de la Recherche (ANR project ``Dirac3D'') under Grant No. ANR-17-CE30-0023.
Furthermore we thank M. Orlita and D. J. Alspaugh for useful discussions.
\end{acknowledgements}

\appendix*
\section{Derive and solve the secular equation for double Dirac quantum well} \label{sec:secular}
We solve Eq. \eqref{eq:schrodinger} for $\Delta(z)$ described by Eq. \eqref{exp:doubledeltazM}. 

If $z<-L/2-l$ or $z>L/2+l$ or $z \in [-\frac{L}{2}+l,\frac{L}{2}-l]$, the equation reads
\begin{equation}
    \label{eq:planewave}
    \partial_z^2 \chi_{\lambda} - K^2 \chi_{\lambda} = 0
\end{equation}
where $K^2 = (1-\omega^2)/\xi^2$. The solutions are a linear combination of $\exp{(Kz)}$ and $\exp{(-Kz)}$. If $z \in [ -\frac{L}{2}-l, -\frac{L}{2} + l ]$, we carry out a change of variable $z+L/2 = \alpha t_L$ and $\alpha^2 = l \xi /2$. The equation then reads
\begin{equation}
    \label{eq:confluentL}
    \partial_{t_L}^2 \chi_{\lambda} - \left( \frac{1}{4} + a_{L,\lambda} \right) \chi_{\lambda} = 0
\end{equation}
where
\begin{equation}
    \label{exp:aL}
     a_{L,\lambda} = - \frac{\lambda}{2} - \frac{l}{2\xi}\omega^2. 
\end{equation}
Eq. \eqref{eq:confluentL} is the standard form of the Weber differential equation whose solution is parabolic cylinder function. By concern for symmetry of the wavefunction, we represent the solution in terms of confluent hypergeometric function $M(a;b;z)$. The even and odd solutions read:
\begin{align}
    \centering
    \label{eq:wavefunctionL}
    u_S (a_{L,\lambda}; t_L) &= e^{-\frac{t_L^2}{4}} M \left(\frac{1}{2}a_{L,\lambda}+\frac{1}{4};\frac{1}{2};\frac{t_L^2}{2} \right) \nonumber \\
    u_A (a_{L,\lambda}; t_L) &= t_L e^{-\frac{t_L^2}{4}} M \left(\frac{1}{2}a_{L,\lambda}+\frac{3}{4};\frac{3}{2};\frac{t_L^2}{2} \right)
\end{align}
where $S$ and $A$ mean symmetric and anti-symmetric, respectively. If $z \in [ \frac{L}{2}-l, \frac{L}{2} + l ]$, we can solve the differential equation and represent the solutions in the similar way. After a change of variable $z-L/2 = \alpha t_R$, 
\begin{equation}
    \label{eq:confluentR}
    \partial_{t_R}^2 \chi_{\lambda} - \left( \frac{1}{4} + a_{R,\lambda} \right) \chi_{\lambda} = 0
\end{equation}
where
\begin{equation}
    \label{exp:aR}
     a_{R,\lambda} =  \frac{\lambda}{2} - \frac{l}{2\xi}\omega^2. 
\end{equation}
Similarly, the solutions for Eq. \eqref{eq:confluentR} are
\begin{align}
    \centering
    \label{eq:wavefunctionR}
    u_S (a_{R,\lambda}; t_R) &= e^{-\frac{t_R^2}{4}} M \left(\frac{1}{2}a_{R,\lambda}+\frac{1}{4};\frac{1}{2};\frac{t_R^2}{2} \right) \nonumber \\
    u_A (a_{R,\lambda}; t_R) &= t_R e^{-\frac{t_R^2}{4}} M \left(\frac{1}{2}a_{R,\lambda}+\frac{3}{4};\frac{3}{2};\frac{t_R^2}{2} \right)
\end{align}
Using the fact that the wavefunction is vanishing at infinity and it is continuous as well as its derivative, we can match the solution in different regions at their common point along the $z-$direction. For simplicity, we note
\begin{align}
    \centering
    \label{exp:notationuv}
    u_{S/A,L/R,\lambda} &= u_{S/A} \left(a_{L/R,\lambda};\sqrt{\frac{2 l}{\xi}} \right) \nonumber \\
    v_{S/A,L/R,\lambda} &=  \frac{\partial}{\partial{t_{L/R}}} u_{S/A} \left(a_{L/R,\lambda}; t_{L/R})\right|_{t_{L/R}=\sqrt{\frac{2 l}{\xi}}}.
\end{align}
Since $\lambda=\pm$ are equivalent when we consider double Dirac QW, we will omit $\lambda$ in the following discussion. The final secular equation reads:
\begin{widetext}
\begin{align}
    \centering
    \label{eq:secular}
    &\left( \sqrt{\frac{l(1-\omega^2)}{\xi}} u_{S,L} + v_{S,L} \right)        \left( \sqrt{\frac{l(1-\omega^2)}{\xi}} u_{A,L} + v_{A,L}  \right)       \left( \sqrt{\frac{l(1-\omega^2)}{\xi}} u_{S,R} + v_{S,R}  \right)          \left( \sqrt{\frac{l(1-\omega^2)}{\xi}} u_{A,R} + v_{A,R}  \right) \nonumber \\
    &= e^{-2 \frac{\sqrt{1-\omega^2}}{\xi}(L-2l)} \left( \frac{l(1-\omega^2)}{\xi} u_{S,R} u_{A,R} - v_{S,R} v_{A,R} \right) \left( \frac{l(1-\omega^2)}{\xi} u_{S,L} u_{A,L} - v_{S,L} v_{A,L} \right)
\end{align}
\end{widetext}

%%%%%%%%%%%%%
\iffalse

\begin{figure}[h]
    \centering
    \subfigure{\includegraphics[width=0.45\textwidth]{chiralwidth.pdf}}
    \subfigure{
    \includegraphics[width= 0.45 \textwidth]{chiralthickness.pdf}
    }
    \caption{Top: $\log_{10} \omega$ vs. $l$ for the $n=0$ chiral state and $L/\xi=20$. Bottom: $\log_{10} \omega$ vs. $L$ for the $n=0$ chiral state and $l/\xi=6$. The results calculated using the secular equation Eq. \eqref{eq:secular} are represented by blue lines and those using Eq. \eqref{eq:splittingsquareB_chiral} by orange lines.}
    \label{fig:widthchiral}
\end{figure}

\fi
%%%%%%%%%%%%%%%%

Let us first try several particular solution to check the validity of our model. Suppose now $\omega=0$ where we know it is impossible for finite $L$ and non-zero $l$. Eq. \eqref{eq:secular} would become
\begin{equation}
    \frac{l}{2\xi} e^{-\frac{2(L-2l)}{\xi}} (\dots)(\dots) = 0
\end{equation}
which cannot be true except when the surface is sharp ($l \ll \xi$) and the distance between two QWs ($L \gg l,\xi$). In fact, when $l \to 0$, there are only three domains along the $z-$direction: $z<-L/2$, $z \in ]-L/2,L/2[$ and $z>L/2$. So we have only two continuity relations for four coefficients, which means two degenerate solutions for $\omega=0$. Another interesting value for $\omega$ is $\omega=1$. We can verify that $\omega=1$ is always a solution of Eq. \eqref{eq:secular} for any parameters. So we also retrieve automatically the bulk spectrum, $E = \pm \sqrt{\hbar^2 v^2 k_{\parallel}^2 + \Delta_0^2}$, in our model.

Next, let's consider the situation when $l \ll \xi$ and derive a formula to evaluate the mass gap of the chiral mode. To do so, we can develop Eq. \eqref{eq:secular} in terms $l/\xi$ and suppose in the first approximation that $\omega$ is at most of same order of $\sqrt{l/\xi}$. After some algebra, we have
\begin{equation}
    2\Delta E =  2\Delta_0 e^{-\frac{L}{\xi}} \sqrt{ 1+ \frac{4l^2}{3\xi^2} } .
\end{equation}

%%%%%%%%%%%%%
\iffalse

\section{Energy splitting of the chiral state} \label{sec:chiralsplitting}
We show the energy splitting of the chiral state when massive VP states emerge.

In Fig. \ref{fig:widthchiral}, we use a scale of $\log_{10}$ for $\omega$ to show the exponential behavior. The top of Fig. \ref{fig:widthchiral} shows the variation of $\omega$ changing $l/\xi$ for the chiral state. We first remark that the energy splitting of the chiral state is $10^4$ times smaller than that of the massive VP states. Our heuristic formula, Eq. \eqref{eq:splittingsquareB_chiral}, gives $10^4$ times bigger than that by the secular equation Eq. \eqref{eq:secular}. The variation of $\omega$ is almost exponential as a function of $l/\xi$. By linear fit, the slope for $\log{\omega}$ vs. $l/\xi$ is approximately 0.9.

The bottom of Fig. \ref{fig:widthchiral} shows the variation of $\omega$ changing $L/\xi$ for $n=0 \dots $ states. Intuitively, the tunneling effect should be exponentially weak when we increase the distance between two QWs. By linear fit, the slope for $\log{\omega}$ vs. $L/\xi$ is precisely -1. This is true for the splitting of the massive states as well as for that of the chiral state. When $L \to \infty$, the splitting would be zero and two splitted levels will merge into one level with the energy of the one single Dirac QW.

\fi
%%%%%%%%%%%%%%%

\bibliography{references}
%%%%%%%%%%%%%%%
\iffalse

%%%%%%%%%% template for picture introduction %%%

\begin{figure}[!h]
  \includegraphics[width=0.5\textwidth]{fig2a}
\caption {Energies of low-lying trion states E$ _{1T} $ and E$ _{2T} $, considering the ground exciton state E$ _{1s} $. Calculations are performed on WS$_{2}$ monolayer using $ \varepsilon_{sub}=2.1, \lambda_{s}=28 $\AA , a) $ \Omega=0 $, b) $ \Omega= $10\AA$ ^{2} $ and $ \Omega= $20\AA$ ^{2} $.}
\label{fig:2a}       % Give a unique label
\end{figure}
\begin{figure}[!h]
  \includegraphics[width=0.5\textwidth]{fig2c}
\caption {Spectrum of a WS$_{2}$ monolayer  showing the effect of Berry correction on exciton X and trion T energies. The parameters used are $ \lambda_{s}=28 $\AA, $ \varepsilon_{sub}=2.1 $ and $ m_{h}=m_{e}=0.34 $.}
\label{fig:2c}       % Give a unique label
\end{figure}

\fi
%%%%%%%%%%%%%%

\end{document}